\documentclass[journal]{IEEEtran}
\usepackage{lineno}
\usepackage{placeins}
\usepackage{times}
\usepackage[pdftex]{graphicx}
\usepackage{lipsum}
\usepackage[cmex10]{amsmath}
\usepackage{amssymb}

\ifCLASSINFOpdf
\else
\fi

\begin{document}
%
\title{Tremor Waveform Denoising and Automatic Location with Neural Network Interpretation}
%
%
%

\author{Claudia Hulbert, Romain Jolivet, Blandine Gardonio, Paul Johnson, Christopher X. Ren, Bertrand Rouet-Leduc
\thanks{C. Hulbert, R. Jolivet and B. Gardonio are with the Laboratoire de G\'eologie, D\'epartment de Geoscience, \'Ecole normale sup\'erieure, PSL University, CNRS UMR 8538, Paris, France. e-mail: (claudia.hulbert@ens.fr).}
\thanks{R. Jolivet is with the Institut Universitaire de France, 1 rue Descartes, 75005 Paris.}
\thanks{B. Rouet-Leduc and P. Johnson and are with the Los Alamos National Laboratory, Geophysics Group, Los Alamos, New Mexico, USA}
\thanks{C. Ren is with the Los Alamos National Laboratory, ISR Group, Los Alamos, New Mexico, USA}%
}

%
%

\markboth{T\MakeLowercase{his work has been submitted to the} IEEE \MakeLowercase{for possible publication}. C\MakeLowercase{opyright may be transferred without notice, after which this version may no longer be accessible.}}%
{Shell \MakeLowercase{\textit{et al.}}: Bare Demo of IEEEtran.cls for Journals}
%



\maketitle

\begin{abstract}
Active faults release tectonic stress imposed by plate motion through a spectrum of slip modes, from slow, aseismic  slip, to dynamic, seismic events. Slow earthquakes are often associated with tectonic tremor, non-impulsive signals that can easily be buried in seismic noise and go undetected.

We present a new methodology aimed at improving the detection and location of tremors hidden within seismic noise. After identifying tremors with a classic convolutional neural network, we rely on neural network attribution to extract core tremor signatures and denoise input waveforms. We then use these cleaned waveforms to locate tremors with standard array-based techniques. 

We apply this method to the Cascadia subduction zone, where we identify tremor patches consistent with existing catalogs. In particular, we show that the cleaned signals resulting from the neural network attribution analysis correspond to a waveform traveling in the Earth's crust and mantle at wavespeeds consistent with local estimates. This approach allows us to extract small signals hidden within the noise, and therefore to locate more tremors than in existing catalogs. 
\end{abstract}


%
\IEEEpeerreviewmaketitle

\section{Introduction}
%
%
%
%

\IEEEPARstart{G}{eodetic} and seismological observations suggest fault slip can take place over a large range of time scales, ranging from seconds or minutes during earthquakes to days, weeks, months or permanently as slow slip \cite{Peng2010, Obara2016, Burgmann2018, Jolivet2020}. Among this spectrum of slip, slow displacements generated by slow slip and creep are often accompanied by characteristic seismic signals termed tremor \cite{Rogers2003,Obara2002,Frank2016,Nadeau2005}. The precise relationship between tremor and slip remains poorly understood; proposed explanations of tremor origins include the breakage of small asperities as the fault displaces, fluid-driven fractures or fluid migration, among others \cite{Seno2003,Ide2007,Cruz2018}. In general tremor has been associated with slow slip, as the source mechanism is consistent with either a double couple mechanism \cite{Shelly2007} or with marked geodetic displacements \cite{Rogers2003}.

Since its discovery in Japan in the early 2000s \cite{Obara2002}, tectonic tremor has been observed in many subduction zones, including Cascadia \cite{Rogers2003, Ide2012}, Alaska \cite{Peterson2009}, Mexico \cite{Payero2008, Ide2012, frank2014using}, Southern Chile \cite{Gallego2013, Ide2012}, New Zealand \cite{Kim2011, Ide2012}, and Costa Rica \cite{Walter2011}. Observation of tremor activity extends to strike-slip fault systems \cite{Nadeau2005, Peng2013,Guo2017}. Like earthquakes, tremor  can  occur spontaneously but can also be triggered, either by tidal loading \cite{Rubinstein2008,Van2016} or by quasistatic or dynamic loading from earthquakes \cite{Peng2013,Gomberg2008}. \\

In contrast to earthquakes that are characterized by impulsive P- and S-wave arrivals, tremor waveforms are not associated with systematic, well-defined patterns. This absence of characteristic traits makes the detection and location of tremors challenging. While identifying an earthquake and picking its phases on a single seismic waveform can be a relatively easy task for an analyst, detecting tremor on a single station is extremely difficult. For this reason, tremor identification typically relies on a large network of seismometers: detecting a coherent signal across the network allows one to discriminate it from noise. 

Methods to locate tremors are usually based on waveform correlation across an array of seismic stations, through the cross-correlation of tremor envelopes from which a differential travel time between stations can be estimated. These differential travel times can then be compared to theoretical ones through grid-search \cite{Obara2002, Nadeau2005}. In some other cases, the entire correlation functions are stacked and its maximum value is used to estimate the associated event location \cite{wech2008, Poiata2016}. 

In addition to envelope-based location, approaches described in the literature include template matching \cite{Shelly2007} and the back-propagation of seismic signals \cite{Kao2005}. Refinements have also been introduced to improve the precision of tremor location, especially regarding depth, which is often poorly resolved. These approaches include double-difference location techniques \cite{Guo2017}, or the estimation of phase lags between P- and S-wave timings \cite{Rocca2009}. \\

Array-based tremor detections and locations are unlikely to work well for low-amplitude tremors, that may appear only at one or two stations and not be identified as coherent signal throughout the array. This type of approach is also particularly challenging in the case of sparse seismic networks, because local sources of noise prohibit high correlations between station pairs. In what follows, we propose a new methodology for tremor location, based upon neural network attribution to extract tremor waveforms. We find that our approach gives results coherent with existing catalogs, and allows the location of tremors below the noise level. This method should prove a valuable tool for the location of small tremors, or in presence of sparse seismic arrays.

 
\section{Method}

\subsection{Denoising tremor waveforms with neural network attribution}

A trained neural network can easily detect tremor at a single seismic station, vastly increasing the number of detections when compared to the catalog it has been trained on \cite{Rouet-Leduc2020}. Our goal here is to leverage the learned representations of a deep neural network to denoise seismic waveforms, revealing the signals that triggered the detection and locating the corresponding tremors. \\

\begin{figure*}[!ht]
\begin{center}
\includegraphics[width=16cm,trim= 0 0 0 0]{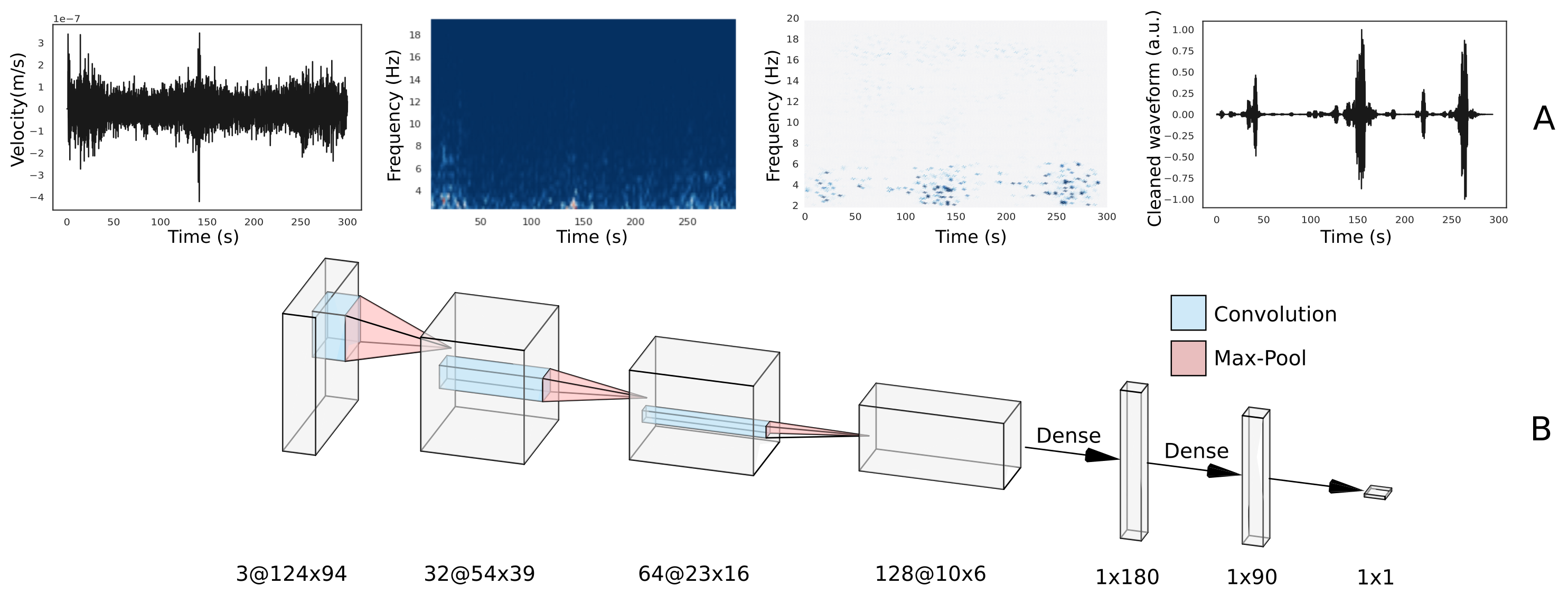}
\caption{\footnotesize{\textbf{Denoising of tremor waveforms with neural network attribution.} Raw waveforms \textbf{(A, left panel)} are turned into spectrograms \textbf{(A, second panel)} and fed to a classic CNN. We rely on a standard network (three convolution+max pool and two dense layers). The architecture of the network is shown in \textbf{(B)}. From these spectrograms, the CNN is tasked to classify tremor from non-tremor. Once the model is trained (with tremors identified from the PNSN catalog), we rely on Integrated Gradients \cite{Sundararajan2017} attribution to interpret its results. \textbf{(A, third panel)} shows the result of the positive attribution for the same waveform; the blue areas correspond to the parts of the spectrogram that carried core tremor information according to the attribution analysis. We can then select exclusively these areas to inverse-Fourier transform, to get a  clean tremor signal \textbf{(A, right panel)}. These cleaned waveforms correspond to the core tremor signals, as seen by our model, and have clear structure compared to the original waveforms; our goal here is to rely on these clean waveforms to locate tremors.}}
\label{fig1}
\end{center}
\end{figure*}

The network we rely upon is a standard convolutional neural network (CNN), with a simple architecture (Figure \ref{fig1} (B)). Our CNN attempts to classify tremor from seismic noise. It is trained on three components, 5-minute spectrograms of approximately 165,000 examples of tremors identified from the PNSN catalog \cite{wech2008}, and 250,000 examples of seismic noise. The noise examples are drawn randomly from times when no tremor has been detected by the PNSN. Examples of noise and tremor are drawn from several seismic stations in Cascadia, located along the Western US coast (nearly all the stations are outside Vancouver Island, our region of interest). The CNN attempts to classify tremor from single, 3-component examples of spectrograms, thereby building a single-station detector. We train the classifier on examples from August 2009 to June 2016. Examples for June 2016 to December 2018 are used for validation and test, with the first half corresponding to the validation set, and the second half to the test set. The performance of our classifier in testing, as measured by the ROC-AUC score, is of 0.913. We leverage adversarial training to further improve the generalization of our model and the interpretability of its gradients \cite{Kim2019}. Details regarding the database used for training, the training and testing procedures, and the performance of the model, can be found in Supplementary Materials. \\

This kind of single station tremor classification tends to generalize well to other areas \cite{Rouet-Leduc2020}: a model trained on waveform spectrograms from a single seismic station in Vancouver Island, Canada, generalizes to other nearby seismic stations, and even to data from Japan - suggesting that tremor waveforms carry some kind of universal signature. 

To probe the classifier for tremor signature patterns, we rely on neural network attribution tools. Neural networks are often considered as black-box models; however there has been considerable effort over the recent years to develop their interpretability. Among these approaches, several attempts at interpretation focused on visualization throughout the hidden layers of the network: i) either by considering all neurons in a given layer as a whole \cite{Yosinski2014}, or ii) by considering each neuron individually, often base on the activation value of each unit \cite{Erhan2009, Zeiler2014, Yosinski2015}. A parallel area of research is to analyze network attribution, \textit{i.e.} the connection between input features and the prediction of the network \cite{Baehrens2010, Simonyan2013, Springenberg2014, Shrikumar2017, Binder2016, Sundararajan2017, Montavon2017}.  \\

We rely on a recent approach for network attribution, using the Integrated Gradients methodology \cite{Sundararajan2017}. Gradients are often used as a basis of attribution techniques, as they can be considered as the coefficients associated to a network's features, and their analysis can therefore inform on important feature contributions. Consider a deep network $N: \mathbb{R}^{n} \rightarrow [0,1]$, and an input datapoint $\textbf{x}=(x_{1},...,x_{n})$. The attribution problem consists in analyzing the contribution of each individual $x_{i}$ to the prediction of the network, $N(\textbf{x})$. For example in image analysis, it would correspond to analyze the contribution of individual pixels to the prediction of the model. 

The Integrated Gradients correspond to a path integral along a straight line connecting a baseline vector to the input vector \textbf{x}:
$$
IG_{i}(\textbf{x})=(x_i-x'_i) \int_{\alpha=0}^{1} \frac{\partial N(\textbf{x}' +\alpha (\textbf{x}-\textbf{x}'))}{\partial x_i} d \alpha,
$$
where $\textbf{x}'$ corresponds to the baseline vector, and $\frac{\partial N(\textbf{x})}{\partial x_i}$ correspond to the gradient of the network along the $i^{th}$ dimension. The baseline vector $\textbf{x}'$ can be taken for example as a black image or a zero embedding vector. Because gradients are integrated in this way, the method is robust to plateaus in the network's response and areas of local gradient instability such as saddle points.\\

The Integrated Gradient approach solves two main issues encountered by several other attribution techniques proposed in the literature: i) it avoids flat, zero gradient regions (plateaus in the output of the network with respect to a region in the input space) that can cause the attribution method to focus on non-relevant features; and ii) functionally equivalent networks (that associate the same inputs to the same outputs) return the same attribution, avoiding the attribution to focus on unimportant characteristics of the classifier. Furthermore this approach is fast to compute. \\

We analyze our input spectrograms with the Integrated Gradients approach, identifying which parts of the spectrograms carry the core of the tremor information. Positive attribution results for a sample waveform is shown in Figure \ref{fig1} (A), third panel. In this plot, the blue areas correspond to the parts of the spectrogram that were crucial for the algorithm to label a waveform as tremor. As expected, the dominant frequency band lies between 2 and 7 Hz, with some smaller contribution from higher frequencies. Note that this analysis is very different from a simple bandpass filter within tremor frequency bands (see Figure \ref{fig2}), because of the temporal distribution of seismic energy in the attribution spectrogram. Focusing exclusively on these areas and performing the associated inverse Fourier transform yields waveforms with clear structure compared to the original signals (Figure \ref{fig1} A) (right-hand panel). It is likely that the attribution analysis identifies some of the individual low-frequency earthquakes (LFEs) contained within the tremor signals. 

\FloatBarrier
\FloatBarrier

\subsection{Reconstructed waveforms are the signature of traveling waves}

\begin{figure}[!ht]
\begin{center}
\includegraphics[width=8.5cm,trim= 0 0 0 0]{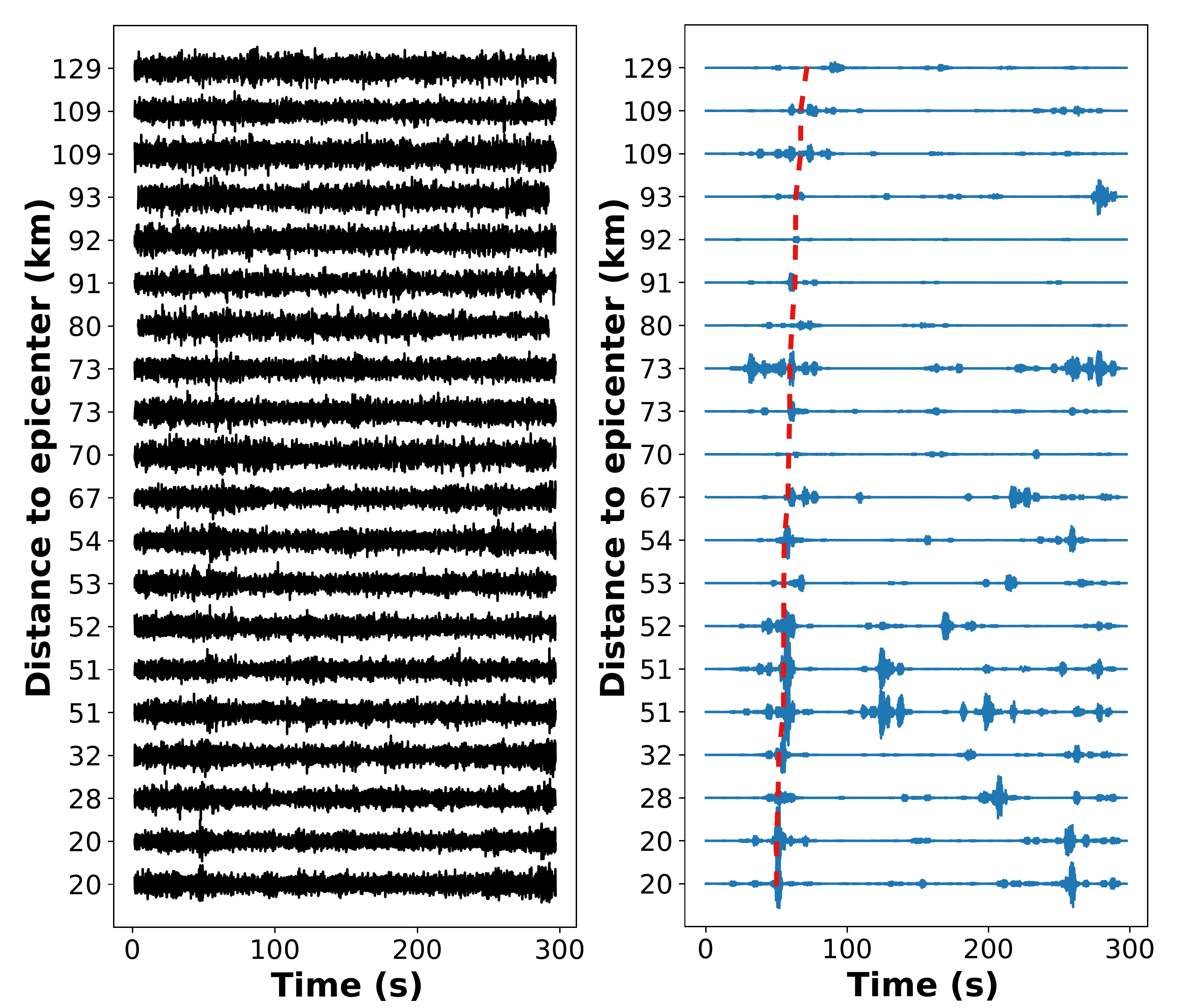}
\caption{\footnotesize{\textbf{Application to an array of seismometers}. \textbf{(A)} Raw waveforms bandpassed between 1 and 8 Hz. \textbf{(B)} Result of the neural network attribution analysis to denoise the waveforms (in blue). The denoising procedure is performed on each station independently. Clean move-out patterns can be seen across the network, which can be compared to shear wave theoretical travel times (red line). The move-outs match theoretical wave speed to first order.}}
\label{fig2a}
\end{center}
\end{figure}

Because seismic energy in the attribution spectrograms is intermittent in time, cleaned waveforms have a much more impulsive structure than the original signals. As illustrated by the example in Figure \ref{fig2a}, raw waveforms (even filtered between 1 and 8Hz) often exhibit weak and emergent signals, which appear much more impulsive once cleaned as described above. Importantly, the reconstructed waveforms show impulsive signals that are coherent between stations.  

We can compare the travel times of the reconstructed signals to the arrival time predictions from a simple 1D velocity model for S-waves. When performing the attribution analysis independently on each station, the observed travel times match the theoretical ones to first order, suggesting envelope cross-correlation location should be possible. This spatio-temporal propagation provides a strong validation that the reconstructed waveforms indeed correspond to signals of seismological interest (here tectonic tremors), extracted from noisy time series. \\

The conservation of travel times in the reconstructed waveforms is also supported by the fact that binary classification results from a CNN are not station dependent. Indeed, applying the same model to other stations, even in different tectonic regions, yields good performance \cite{Rouet-Leduc2020}. Therefore we expect that tremor signatures are shared across stations, and, if detected, these should consequentially align across a seismic network according to true move-out patterns. \\

\FloatBarrier
\FloatBarrier

\subsection{Location procedure}

\begin{figure}[!ht]
\begin{center}
\includegraphics[width=8.5cm,trim= 0 0 0 0]{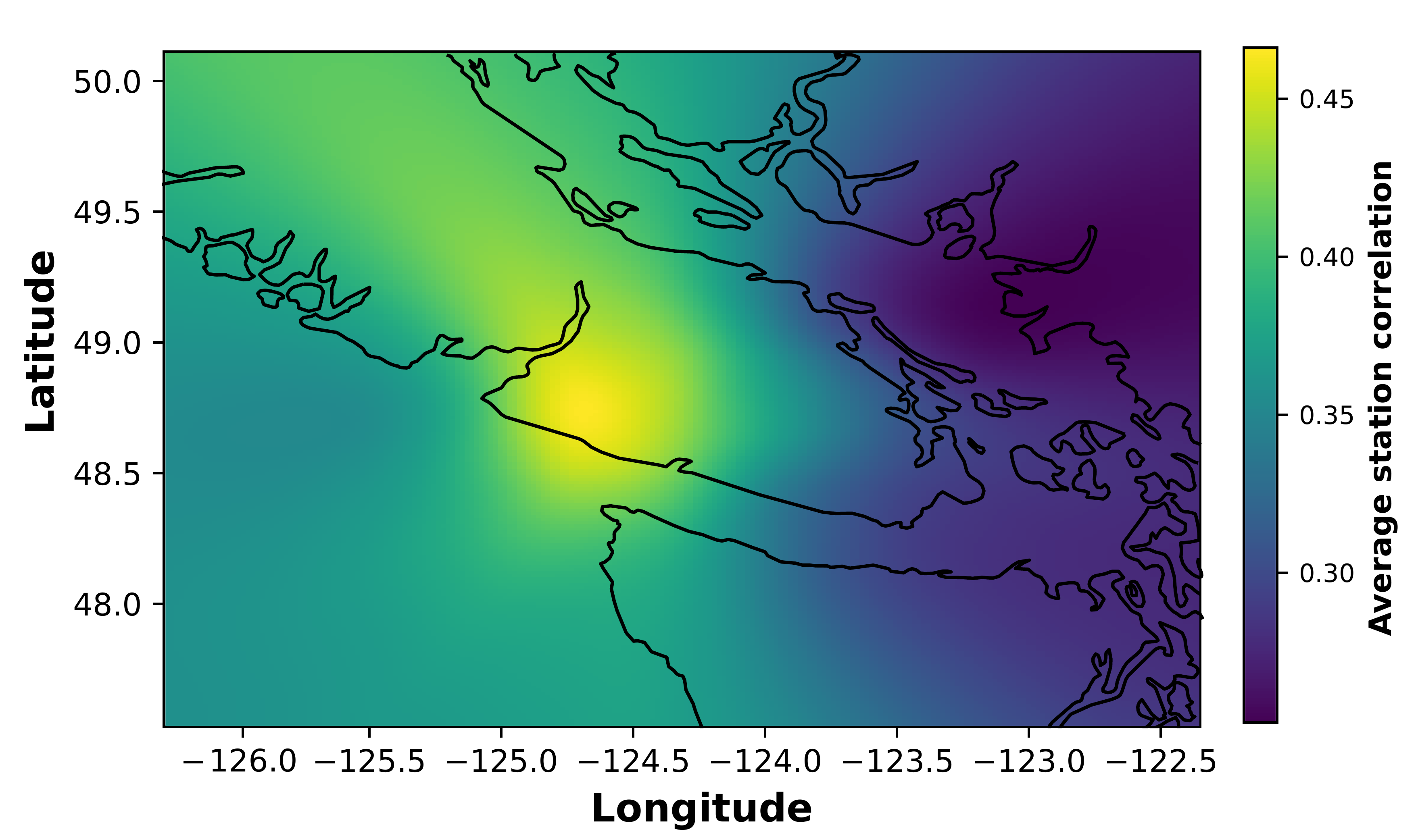}
\caption{\footnotesize{\textbf{Example tremor location.} We compute the envelope of the cleaned signals; these envelopes are then sliced around each identified STA/LTA pick, and cross-correlated for all station pairs. This figure shows the stacked correlation plotted over the lat/lon grid for one tremor. If a number of criteria are met (see below), the maximum of the stacked correlation function is returned as tremor location.}}
\label{fig2}
\end{center}
\end{figure}

If indeed our reconstructed waveforms preserve the tremors' move-out patterns, it should be possible to use them to locate. Once all waveforms of interest are reconstructed as described above, we rely on traditional array-based methods to locate tremors. Our approach builds upon the afore-mentioned hypothesis that denoising waveforms at different stations will result in cleaned waveforms that preserve the tremors' move-out structure, as shown on the example in Figure \ref{fig2}.

We rely here on a standard array-based methodology to locate tremors. We begin by running a classic STA/LTA picker through the cleaned waveforms. The fact that these waveforms are more impulsive and structured than the original signals makes the picking procedure more effective, resulting in many picks identified. These waveforms are then sliced around each pick, using a window size based upon the maximum theoretical travel time that can be observed across the array. \\

We then measure the correlation of the envelopes of the sliced waveforms, for each pair of stations within the array. The envelopes are computed as the smoothed RMS of the denoised waveforms, in a way similar to \cite{Nakamura2017}. Each envelope correlation value is associated to a time lag between the two station waveforms, which can be compared with shear wave theoretical time differences between station pairs. A regional 1D velocity model is used to estimate travel times \cite{Crosson1978}. For each station, both horizontal components are used for the analysis. We keep the entire correlation functions to locate the events \cite{wech2008,Poiata2016}; these correlation functions are stacked for all pairs of stations that display a high envelope correlation (see below). The point in the spatial grid corresponding to the maximum stacked correlation is then returned as source location for one particular tremor. 

Figure \ref{fig2} shows a stacked correlation function over our region of interest. The maximum of the stacked correlation corresponds to the tremor's location. Because we rely on envelope cross-correlation, depth is not well constrained. Note that if the reconstructed waveforms were not consistent with the speed of S-wave in Earth, the stacked correlation would not coalesce to a maximum, as it does for example on Figure 3.\\ 

\FloatBarrier
\FloatBarrier

We use several selection criteria to improve confidence in a tremor's location: 

i) We only locate the source of waveforms that the neural network model classifies as tremor with a `confidence' threshold (softmax $ \geq 0.7$, on a minimum of two stations).

ii) To focus the analysis on informative stations, a station pair is only considered if its maximum correlation value reaches a given threshold. In line with the literature, we set a minimum correlation threshold of 0.7 \cite{wech2008, Nakamura2017}. At least 5 station pairs must reach this minimum threshold. The stacked correlation values are weighted according to the maximum correlation measured for each station pair - as station pairs with higher waveform correlations are likely to be closer to the event source, or to exhibit a higher signal-to-noise ratio.

iii) To ensure that the stacked correlation maximum is well defined, we compute the ratio between the maximum stacked correlation value and the median absolute deviation (MAD) of the correlation function. For tremor identification, in line with \cite{Shelly2007} we only keep events with a MAD ratio above 8 and discard the others.

iv) To limit the locations' spatial uncertainty, we use an approach similar to \cite{wech2008}. We randomly remove a subset of 10\% of the stations pairs by bootstrapping, over a number of iterations. We discard signals for which the bootstrapped locations are not consistent spatially (outside a 5km radius around the initial location). 

\section{Results}

\subsection{Area analyzed}

To test the performance of our location methodology, we apply it to the 2018 Cascadia slow slip. We focus on Southern Victoria Island, Canada, where clear tremor activity is typically recorded \cite{Rogers2003, wech2008}. In this region the Cascadia subduction zone experiences an ETS approximately every 14 months, which typically lasts for one or two months. According to the PNSN tremor logs, the 2018 event started in early May, and paused for a while before resuming in mid-June until mid-July 2018. We therefore analyze three months of data (May, June and July 2018). Applying our method to a well-studied area allows us to validate it by comparing our results to existing tremor catalogs. \\

\begin{figure}[!ht]
\begin{center}
\includegraphics[width=8.5cm,trim= 0 0 0 0]{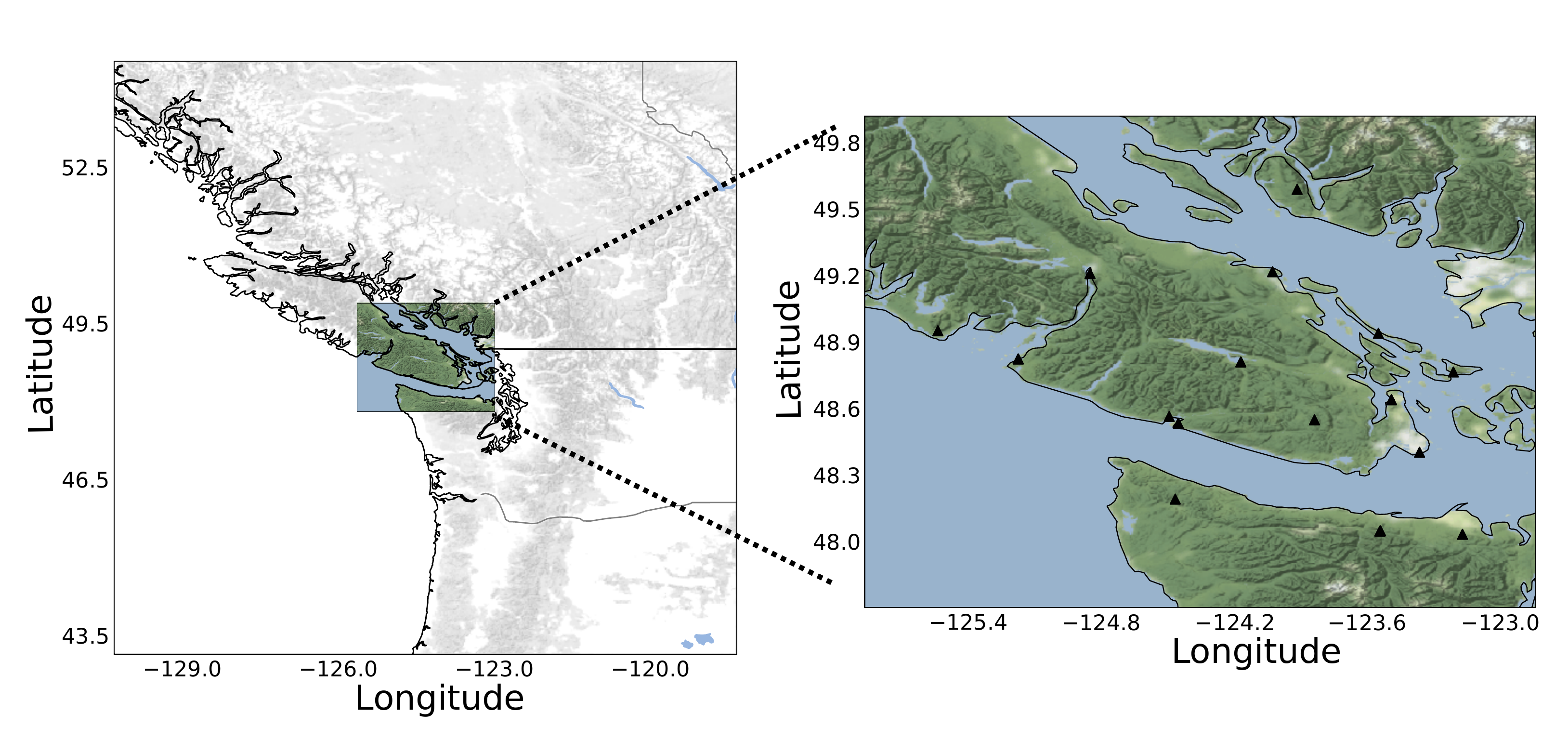}
\caption{\footnotesize{\textbf{Area analyzed and seismic array used.} Left: Map of the Southern Vancouver Island, Canada, analyzed in this study. Right: seismic array. The black triangles correspond to seismic stations used to locate tremors.} }
\label{fig3a}
\end{center}
\end{figure}

We rely on an array of 21 seismometers for locating the tremors, from the CNSN \cite{FDSN} and PBO \cite{PBO} seismic networks. About a third of the stations are borehole instruments. Figure \ref{fig3a} shows a map of the area analyzed, as well as the seismic array used.

\FloatBarrier
\begin{figure*}[!ht]
\begin{center}
\includegraphics[width=18cm,trim= 0 0 0 0]{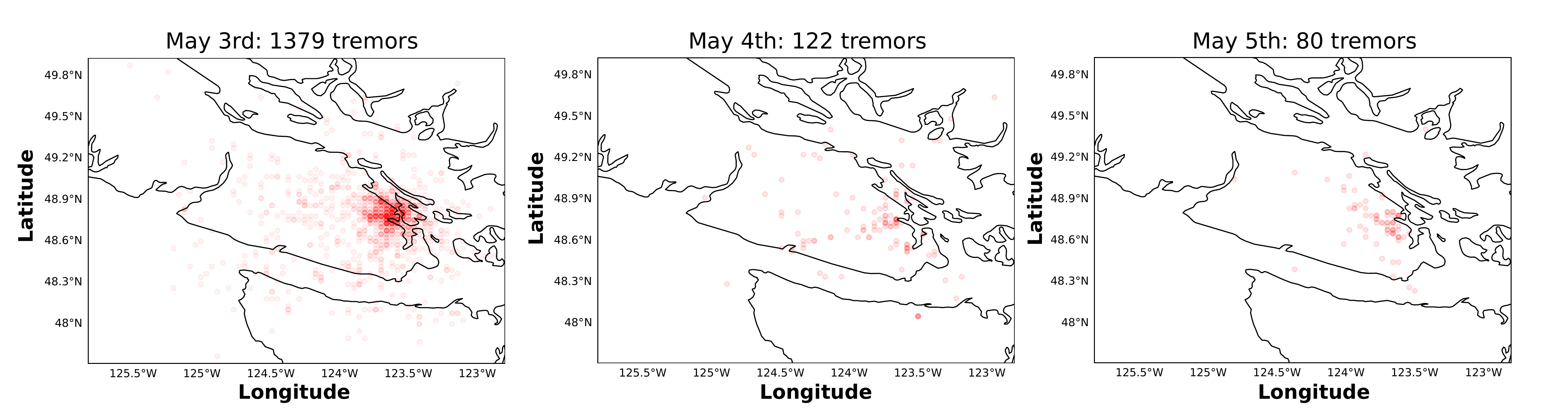}
\caption{\footnotesize{\textbf{Example location.} Tremor locations (in red) for three consecutive days with consistent, local tremor patches, during the 2018 slow slip event.}} 
\label{fig3b}
\end{center}
\end{figure*}
\FloatBarrier

\subsection{Application to the 2018 Cascadia ETS}

In Figure \ref{fig3b} we illustrate our results by showing tremor locations for three consecutive days during the 2018 SSE. For comparison, the PNSN reported 223 tremors in this patch on May 3rd, 0 on May 4th, and 7 on May 5th. We chose to display these three days in particular because they contain consistent tremor patches, located around the same area. The fact that we identify consistent tremor activity in the same area over the three days suggests that our additional detections are likely to be accurate. Furthermore, local tremor patches as shown here are a better test to validate our methodology, as they are less forgiving regarding the location of false positives, or mis-locations of real events. Most of the outliers to the main patches are relatively close to each other, which suggests that they may correspond to smaller, weaker tremor patches. Individual tremors that are far from each other may correspond to mis-located signals or false positives. 

Overall, we find our locations to be consistent with tremor locations in existing catalogs \cite{wech2008}, even for days characterised by small, local patches such as those displayed above. Figures S1, S2 and S3 of the Supplementary show the distribution of tremor locations for all days analyzed, with comparisons to locations from the PNSN catalog. \\

\begin{figure}[!ht]
\begin{center}
\includegraphics[width=9cm,trim= 0 0 0 0]{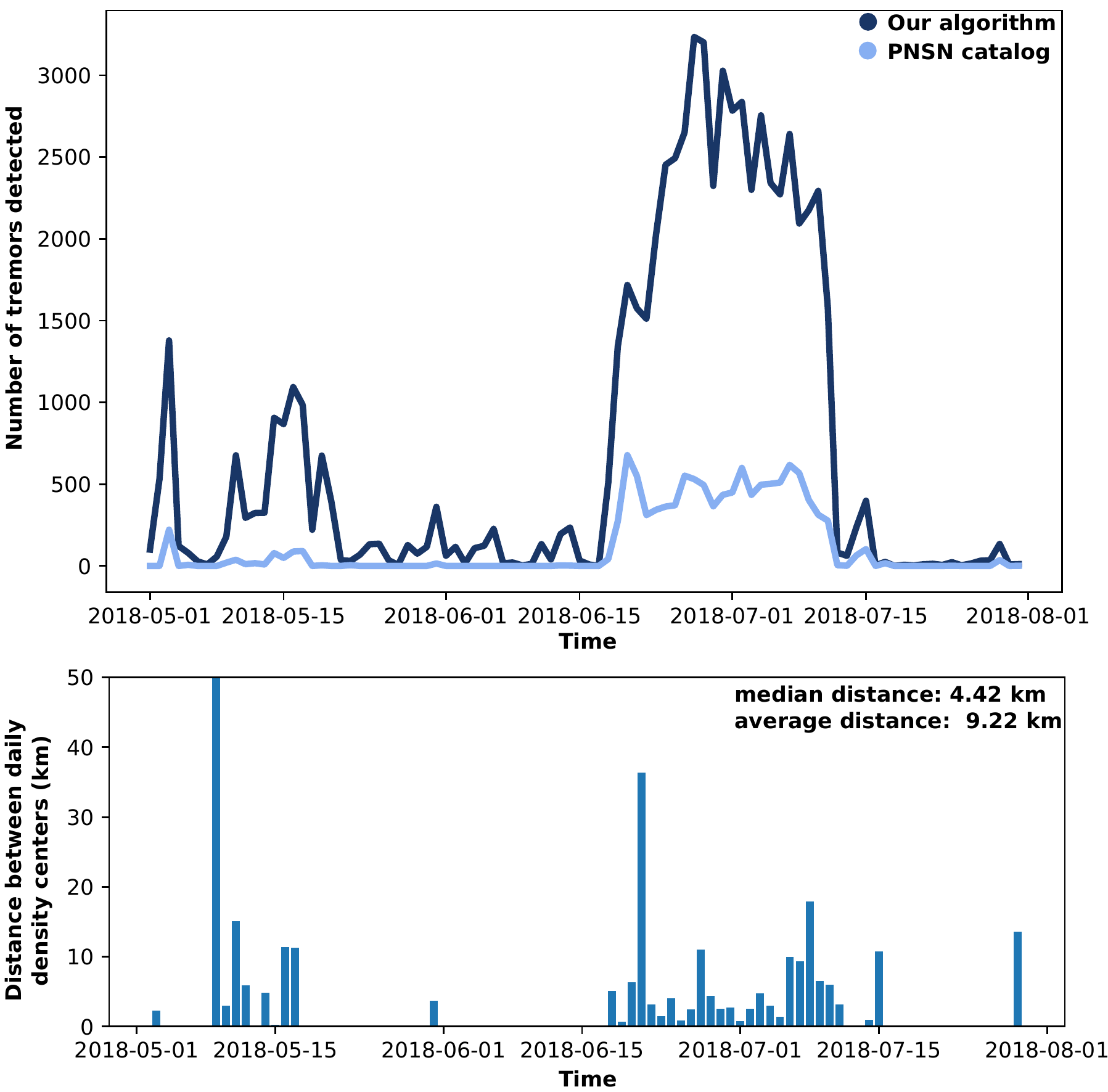}
\caption{\footnotesize{\textbf{Comparison with the PNSN catalog.} \textbf{(A)} Number of tremors detected by our algorithm (dark blue), compared to the number of detected tremors in the PNSN catalog (light blue). The timing of detections is consistent between both catalogs, with a high number of tremors detected between June 18th and July 11th. \textbf{(B)} Distance between daily tremor patches detected by our algorithm and by the PNSN catalog. For most days, the patches have similar locations (with a median distance of 4.42km). The two days where the distance is high correspond to patches in the PNSN catalog that are on the border or outside of our seismic array.}}
\label{fig4}
\end{center}
\end{figure}

Figure \ref{fig4} summarizes the comparison between our results and the tremor locations from the PNSN catalog for the 2018 SSE. In particular, Figure \ref{fig4} \textbf{(A)} shows the number of tremors detected by both approaches, for all days during the analysis. Variations in the number of detections are consistent, with a higher number of tremors detected between the 18th of June and the 11th of July. Overall our algorithm identifies 6 times more tremors than the PNSN catalog (67000 vs 11000 tremors over the period and the region analyzed). Part of this increase in detection numbers can be attributed to methodological differences between the approaches: the PNSN classifies 30s of waveforms as tremor, whereas our algorithm looks at individual picks in the waveforms, and can therefore go below 30s. However, part of the additional detections are likely to come from the ability of our methodology to capture small signals below the noise level. This hypothesis is supported by the fact that for many days, such as the 4th of May shown above, our methodology finds a relatively high number of tremors (more than a hundred), while the PNSN reports no tremor activity. Note that both algorithms use the same correlation threshold as criterion for considering station pairs (0.7), and therefore the higher number of located tremors in our analysis is not the consequence of using a smaller threshold.

Figure \ref{fig4} \textbf{(B)} shows how close the identified tremor patches in both catalogs are, on a daily basis. We compute the density center of daily tremor distribution, for all days with tremor activity underneath Vancouver Island according to the PNSN catalog (when 10 or more tremors are detected). We then compute how far our daily tremor density centers are compared to the tremor density centers reported in the PNSN catalog. For a few days characterized by several tremor patches, we rely on spectral clustering to separate the patches, and compare individual patches (see Supplementary).  \\

\FloatBarrier
\FloatBarrier

For most days, distances between tremor patches identified in both catalogs are small, on the order of a few km. The median distance over the period analyzed is 4.42km (less than one grid point), showing that indeed daily tremors are identified at very similar locations. However, for two particular days (9th of May and 21st of June), the distances are large, on the order of several dozens of km. These two days correspond to times when identified tremor patches in the PNSN catalog are small and located in the North-West and the South-East, at the edge or outside of our seismic array. This makes location difficult for our algorithm, as fewer stations are likely to see the tremors and reach a high correlation value. To overcome this issue in the future, we will work on locating events with overlapping sub-arrays, which should alleviate the issue associated with the borders of the network. On one day (July 17th), our algorithm also misses a tremor patch identified by the PNSN. As there are not enough datapoints in our catalog to compute a density center for July 17th, compared locations for this day can be found in the Supplementary.

\section{Conclusion}

In contrast to earthquakes, tremor waveforms are emergent which makes arrivals are extremely difficult to identify and pick. Therefore, traditional location techniques tend to break down, and coherent envelope signals observed across an array are often used instead for location. This makes tremor location particularly vulnerable to local sources of noise, that can impact drastically the correlation values observed between station pairs. Furthermore, for the same reason sparse arrays are likely to be able to locate only tremors of high amplitude, that would remain correlated across the seismic network. \\

We developed a new tremor location approach, that is more robust to seismic noise. We show that waveform reconstruction through neural network attribution is a powerful tool for the detection and location of small tremors. We find that the reconstructed waveforms are the signatures of travelling waves, consistent with S-waves theoretical travel times. Our methodology is validated by the analysis of the 2018 SSE in Vancouver Island. \\  

This methodology will be particularly helpful in areas with sparse seismic arrays, for which local noise sources are likely to mask signals of interest and prevent the successful cross-correlation of tremor envelopes. Finally, our approach is fully automatic and can be deployed as a systematic tool for tremor detection and location with minimal human intervention.


%


\section*{Acknowledgment}
We thank I. McBrearty for useful discussions. The seismic data used were obtained from the Canadian National Seismic Network \cite{FDSN} and the Plate Boundary Observatory \cite{PBO}. All the data is publicly available. 

C. H. and B. G. were supported by the joint research laboratory effort in the framework of the CEA-ENS Yves Rocard LRC (France). Both CH and R. J. were supported by the European Research Council (ERC) under the European Union’s Horizon 2020 research and innovation program (Geo-4D project, grant agreement 758210). B. R.L.'s work was funded by Institutional Support (LDRD) at Los Alamos (20200278ER). P. J. was supported by DOE Office of Science (Geoscience Program, grant 89233218CNA000001), and C. R. by .

CH developed the automatic location approach. RJ supervised the project and helped interpreting the results.  CH and BRL developed the neural network. BG helped analyzing the waveforms. All authors contributed to writing the manuscript.

\ifCLASSOPTIONcaptionsoff
  \newpage
\fi



\bibliographystyle{IEEEtran}

\begin{thebibliography}{10}
\providecommand{\url}[1]{#1}
\csname url@samestyle\endcsname
\providecommand{\newblock}{\relax}
\providecommand{\bibinfo}[2]{#2}
\providecommand{\BIBentrySTDinterwordspacing}{\spaceskip=0pt\relax}
\providecommand{\BIBentryALTinterwordstretchfactor}{4}
\providecommand{\BIBentryALTinterwordspacing}{\spaceskip=\fontdimen2\font plus
\BIBentryALTinterwordstretchfactor\fontdimen3\font minus
  \fontdimen4\font\relax}
\providecommand{\BIBforeignlanguage}[2]{{%
\expandafter\ifx\csname l@#1\endcsname\relax
\typeout{** WARNING: IEEEtran.bst: No hyphenation pattern has been}%
\typeout{** loaded for the language `#1'. Using the pattern for}%
\typeout{** the default language instead.}%
\else
\language=\csname l@#1\endcsname
\fi
#2}}
\providecommand{\BIBdecl}{\relax}
\BIBdecl

\bibitem{Peng2010}
Z.~Peng and J.~Gomberg, ``An integrated perspective of the continuum between
  earthquakes and slow-slip phenomena,'' \emph{Nature Geoscience}, vol.~3,
  no.~9, p. 599, 2010.

\bibitem{Obara2016}
\BIBentryALTinterwordspacing
K.~Obara and A.~Kato, ``Connecting slow earthquakes to huge earthquakes,''
  \emph{Science}, vol. 353, no. 6296, pp. 253--257, 2016. [Online]. Available:
  \url{http://science.sciencemag.org/content/353/6296/253}
\BIBentrySTDinterwordspacing

\bibitem{Burgmann2018}
R.~Bürgmann, ``The geophysics, geology and mechanics of slow fault slip,''
  \emph{Earth and Planetary Science Letters}, vol. 495, pp. 112 -- 134, 2018.

\bibitem{Jolivet2020}
R.~Jolivet and W.~B. Frank, ``The transient and intermittent nature of slow
  slip,'' \emph{AGU Advances}, vol.~1, no.~1, p. e2019AV000126, 2020.

\bibitem{Rogers2003}
G.~Rogers and H.~Dragert, ``Episodic tremor and slip on the cascadia subduction
  zone: The chatter of silent slip,'' \emph{Science}, vol. 300, no. 5627, pp.
  1942--1943, 2003.

\bibitem{Obara2002}
K.~Obara, ``Nonvolcanic deep tremor associated with subduction in southwest
  japan,'' \emph{Science}, vol. 296, no. 5573, pp. 1679--1681, 2002.

\bibitem{Frank2016}
W.~B. Frank, ``Slow slip hidden in the noise: The intermittence of tectonic
  release,'' \emph{Geophysical Research Letters}, vol.~43, no.~19, pp.
  10,125--10,133, 2016.

\bibitem{Nadeau2005}
R.~M. Nadeau and D.~Dolenc, ``Nonvolcanic tremors deep beneath the san andreas
  fault,'' \emph{Science}, vol. 307, no. 5708, pp. 389--389, 2005.

\bibitem{Seno2003}
T.~Seno and T.~Yamasaki, ``Low-frequency tremors, intraslab and interplate
  earthquakes in southwest japan—from a viewpoint of slab dehydration,''
  \emph{Geophysical Research Letters}, vol.~30, no.~22, 2003.

\bibitem{Ide2007}
S.~Ide, G.~C. Beroza, D.~R. Shelly, and T.~Uchide, ``A scaling law for slow
  earthquakes,'' \emph{Nature}, vol. 447, no. 7140, p.~76, 2007.

\bibitem{Cruz2018}
V.~M. Cruz-Atienza, C.~Villafuerte, and H.~S. Bhat, ``Rapid tremor migration
  and pore-pressure waves in subduction zones,'' \emph{Nature communications},
  vol.~9, no.~1, pp. 1--13, 2018.

\bibitem{Shelly2007}
D.~R. Shelly, G.~C. Beroza, and S.~Ide, ``Non-volcanic tremor and low-frequency
  earthquake swarms,'' \emph{Nature}, vol. 446, no. 7133, p. 305, 2007.

\bibitem{Ide2012}
S.~Ide, ``Variety and spatial heterogeneity of tectonic tremor worldwide,''
  \emph{Journal of Geophysical Research: Solid Earth}, vol. 117, no.~B3, 2012.

\bibitem{Peterson2009}
C.~L. Peterson and D.~H. Christensen, ``Possible relationship between
  nonvolcanic tremor and the 1998–2001 slow slip event, south central
  alaska,'' \emph{Journal of Geophysical Research: Solid Earth}, vol. 114,
  no.~B6, 2009.

\bibitem{Payero2008}
J.~S. Payero, V.~Kostoglodov, N.~Shapiro, T.~Mikumo, A.~Iglesias,
  X.~Pérez-Campos, and R.~W. Clayton, ``Nonvolcanic tremor observed in the
  mexican subduction zone,'' \emph{Geophysical Research Letters}, vol.~35,
  no.~7, 2008.

\bibitem{frank2014using}
W.~B. Frank, N.~M. Shapiro, A.~L. Husker, V.~Kostoglodov, A.~Romanenko, and
  M.~Campillo, ``Using systematically characterized low-frequency earthquakes
  as a fault probe in guerrero, mexico,'' \emph{Journal of Geophysical
  Research: Solid Earth}, vol. 119, no.~10, pp. 7686--7700, 2014.

\bibitem{Gallego2013}
A.~Gallego, R.~M. Russo, D.~Comte, V.~Mocanu, R.~E. Murdie, and J.~VanDecar,
  ``Tidal modulation of continuous nonvolcanic seismic tremor in the chile
  triple junction region,'' \emph{Geochemistry, Geophysics, Geosystems},
  vol.~14, no.~4, pp. 851--863, 2013.

\bibitem{Kim2011}
M.~J. Kim, S.~Y. Schwartz, and S.~Bannister, ``Non-volcanic tremor associated
  with the march 2010 gisborne slow slip event at the hikurangi subduction
  margin, new zealand,'' \emph{Geophysical Research Letters}, vol.~38, no.~14,
  2011.

\bibitem{Walter2011}
J.~I. Walter, S.~Y. Schwartz, J.~M. Protti, and V.~Gonzalez, ``Persistent
  tremor within the northern costa rica seismogenic zone,'' \emph{Geophysical
  Research Letters}, vol.~38, no.~1, 2011.

\bibitem{Peng2013}
Z.~Peng, H.~Gonzalez-Huizar, K.~Chao, C.~Aiken, B.~Moreno, and G.~Armstrong,
  ``Tectonic tremor beneath cuba triggered by the m w 8.8 maule and m w 9.0
  tohoku-oki earthquakes,'' \emph{Bulletin of the Seismological Society of
  America}, vol. 103, no.~1, pp. 595--600, 2013.

\bibitem{Guo2017}
H.~Guo, H.~Zhang, R.~M. Nadeau, and Z.~Peng, ``High-resolution deep tectonic
  tremor locations beneath the san andreas fault near cholame, california,
  using the double-pair double-difference location method,'' \emph{Journal of
  Geophysical Research: Solid Earth}, vol. 122, no.~4, pp. 3062--3075, 2017.

\bibitem{Rubinstein2008}
J.~L. Rubinstein, M.~La~Rocca, J.~E. Vidale, K.~C. Creager, and A.~G. Wech,
  ``Tidal modulation of nonvolcanic tremor,'' \emph{Science}, vol. 319, no.
  5860, pp. 186--189, 2008.

\bibitem{Van2016}
N.~J. van~der Elst, A.~A. Delorey, D.~R. Shelly, and P.~A. Johnson,
  ``Fortnightly modulation of san andreas tremor and low-frequency
  earthquakes,'' \emph{Proceedings of the National Academy of Sciences}, vol.
  113, no.~31, pp. 8601--8605, 2016.

\bibitem{Gomberg2008}
J.~Gomberg, J.~L. Rubinstein, Z.~Peng, K.~C. Creager, J.~E. Vidale, and
  P.~Bodin, ``Widespread triggering of nonvolcanic tremor in california,''
  \emph{Science}, vol. 319, no. 5860, pp. 173--173, 2008.

\bibitem{wech2008}
A.~G. Wech and K.~C. Creager, ``{Automated detection and location of Cascadia
  tremor},'' \emph{Geophysical Research Letters}, vol.~35, no.~20, 2008.

\bibitem{Poiata2016}
N.~Poiata, C.~Satriano, J.-P. Vilotte, P.~Bernard, and K.~Obara, ``Multiband
  array detection and location of seismic sources recorded by dense seismic
  networks,'' \emph{Geophysical Journal International}, vol. 205, no.~3, pp.
  1548--1573, 2016.

\bibitem{Kao2005}
H.~Kao, S.-J. Shan, H.~Dragert, G.~Rogers, J.~F. Cassidy, and K.~Ramachandran,
  ``A wide depth distribution of seismic tremors along the northern cascadia
  margin,'' \emph{Nature}, vol. 436, no. 7052, p. 841, 2005.

\bibitem{Rocca2009}
M.~La~Rocca, K.~C. Creager, D.~Galluzzo, S.~Malone, J.~E. Vidale, J.~R. Sweet,
  and A.~G. Wech, ``Cascadia tremor located near plate interface constrained by
  s minus p wave times,'' \emph{Science}, vol. 323, no. 5914, pp. 620--623,
  2009.

\bibitem{Rouet-Leduc2020}
B.~Rouet-Leduc, C.~Hulbert, I.~W. McBrearty, and P.~A. Johnson, ``Probing slow
  earthquakes with deep learning,'' \emph{Geophysical Research Letters},
  vol.~47, no.~4, p. e2019GL085870, 2020.

\bibitem{Kim2019}
B.~Kim, J.~Seo, and T.~Jeon, ``Bridging adversarial robustness and gradient
  interpretability,'' \emph{arXiv preprint arXiv:1903.11626}, 2019.

\bibitem{Yosinski2014}
J.~Yosinski, J.~Clune, Y.~Bengio, and H.~Lipson, ``How transferable are
  features in deep neural networks?'' in \emph{Advances in neural information
  processing systems}, 2014, pp. 3320--3328.

\bibitem{Erhan2009}
D.~Erhan, Y.~Bengio, A.~Courville, and P.~Vincent, ``Visualizing higher-layer
  features of a deep network,'' \emph{University of Montreal}, vol. 1341,
  no.~3, p.~1, 2009.

\bibitem{Zeiler2014}
M.~D. Zeiler and R.~Fergus, ``Visualizing and understanding convolutional
  networks,'' in \emph{European conference on computer vision}.\hskip 1em plus
  0.5em minus 0.4em\relax Springer, 2014, pp. 818--833.

\bibitem{Yosinski2015}
J.~Yosinski, J.~Clune, A.~Nguyen, T.~Fuchs, and H.~Lipson, ``Understanding
  neural networks through deep visualization,'' \emph{arXiv preprint
  arXiv:1506.06579}, 2015.

\bibitem{Baehrens2010}
D.~Baehrens, T.~Schroeter, S.~Harmeling, M.~Kawanabe, K.~Hansen, and K.-R.
  M{\"u}ller, ``How to explain individual classification decisions,'' \emph{The
  Journal of Machine Learning Research}, vol.~11, pp. 1803--1831, 2010.

\bibitem{Simonyan2013}
K.~Simonyan, A.~Vedaldi, and A.~Zisserman, ``Deep inside convolutional
  networks: Visualising image classification models and saliency maps,''
  \emph{arXiv preprint arXiv:1312.6034}, 2013.

\bibitem{Springenberg2014}
J.~T. Springenberg, A.~Dosovitskiy, T.~Brox, and M.~Riedmiller, ``Striving for
  simplicity: The all convolutional net,'' \emph{arXiv preprint
  arXiv:1412.6806}, 2014.

\bibitem{Shrikumar2017}
A.~Shrikumar, P.~Greenside, and A.~Kundaje, ``Learning important features
  through propagating activation differences,'' \emph{arXiv preprint
  arXiv:1704.02685}, 2017.

\bibitem{Binder2016}
A.~Binder, G.~Montavon, S.~Lapuschkin, K.-R. M{\"u}ller, and W.~Samek,
  ``Layer-wise relevance propagation for neural networks with local
  renormalization layers,'' in \emph{International Conference on Artificial
  Neural Networks}.\hskip 1em plus 0.5em minus 0.4em\relax Springer, 2016, pp.
  63--71.

\bibitem{Sundararajan2017}
M.~Sundararajan, A.~Taly, and Q.~Yan, ``Axiomatic attribution for deep
  networks,'' \emph{arXiv preprint arXiv:1703.01365}, 2017.

\bibitem{Montavon2017}
\BIBentryALTinterwordspacing
G.~Montavon, S.~Lapuschkin, A.~Binder, W.~Samek, and K.-R. Müller,
  ``Explaining nonlinear classification decisions with deep taylor
  decomposition,'' \emph{Pattern Recognition}, vol.~65, pp. 211 -- 222, 2017.
  [Online]. Available:
  \url{http://www.sciencedirect.com/science/article/pii/S0031320316303582}
\BIBentrySTDinterwordspacing

\bibitem{Nakamura2017}
M.~Nakamura, ``Distribution of low-frequency earthquakes accompanying the very
  low frequency earthquakes along the ryukyu trench,'' \emph{Earth, Planets and
  Space}, vol.~69, no.~1, p.~49, 2017.

\bibitem{Crosson1978}
R.~S. Crosson and L.~J. Noson, \emph{Compilation of Earthquake Hypocenters in
  Western Washington-1975}.\hskip 1em plus 0.5em minus 0.4em\relax Department
  of Natural Resources, 1978, vol.~64.

\bibitem{FDSN}
``Canadian national seismograph network,'' \emph{Geological Survey of Canada},
  1989.

\bibitem{PBO}
P.~G. Silver, Y.~Bock, D.~C. Agnew, T.~Henyey, A.~T. Linde, T.~V. McEvilly,
  J.-B. Minster, B.~A. Romanowicz, I.~Sachs, R.~B. Smith \emph{et~al.}, ``A
  plate boundary observatory,'' \emph{Iris Newsletter}, vol.~16, no.~2, p.~3,
  1999.

\end{thebibliography}
\end{document}